\documentclass[traditabstract]{aa}

\ifx\pdfoutput\undefined
\usepackage{graphicx}
\else
\usepackage[pdftex]{graphicx}
\usepackage[pdftex,colorlinks,citecolor=blue,linkcolor=blue,pdfauthor= Matteo~Maturi,pdftitle =Filament~detection,pdfsubject=Cosmology]{hyperref}
\usepackage{epstopdf}
\fi

\usepackage{txfonts}
\usepackage{natbib}
\usepackage{hyperref}

\renewcommand{\d}{\mathrm{d}}

\newcommand{\be}{\begin{equation}}
\newcommand{\ee}{\end{equation}}

\def\be{\begin{equation}}
\def\ee{\end{equation}}

\begin{document} 

\title{Weak lensing detection of intra-cluster filaments\\ with ground based data}

\author{Matteo Maturi\inst{1}\thanks{\email{maturi@uni-heidelberg.de}}
    \and Julian Merten\inst{2}\thanks{\email{jmerten@caltech.edu}}}

\titlerunning{}
\authorrunning{M. Maturi and J. Merten}

\institute{
  $^1$~Zentrum f\"ur Astronomie der Universit\"at Heidelberg, Institut
  f\"ur Theoretische Astrophysik, Philosophenweg 12, 69120
  Heidelberg, Germany\\
  $^2$~Jet Propulsion Laboratory, California Institute of Technology, 4800 Oak Grove Drive,
  MS 169-237, Pasadena, CA 91109, USA
}

\date{\emph{Astronomy \& Astrophysics, submitted}}

\abstract{
According to the current standard model of Cosmology, matter in the Universe arranges itself along a network of 
filamentary structure. These filaments connect the main nodes of this so-called 'Cosmic Web', which are clusters of galaxies. 
Although its large-scale distribution is clearly characterized by numerical simulations, constraining the dark matter content of 
the cosmic web in reality turns out to be difficult. The natural method of choice is gravitational lensing.
%, 
%which has already proven the ability to map the distribution of matter in galaxy clusters to a great level of detail. 
However, the direct detection and mapping of the elusive filament signal is challenging and in this work we present two methods,
specifically tailored to achieve this task.

A linear matched filter aims at the detection of the smooth mass
component of filaments and is optimized to perform a shear
decomposition that follows the anisotropic component of the lensing
signal. Filaments clearly inherit this property due to their
morphology. At the same time, the contamination arising from the
central massive cluster is controlled in a natural way. The filament
$1 \sigma$ detection is of about $\kappa \sim 0.01-0.005$ according to the
filter's template width and length, enabling the detection of
structures out of reach with other approaches.

The second, complementary method seeks to detect the clumpy component of filaments. The detection is 
determined by the number density of sub-clump identifications in an area enclosing the potential filament, 
as it was found within the observed field with the filter approach.

We test both methods against Mock observations based on realistic
N-Body simulations of filamentary structure and prove the feasibility
of detecting filaments with ground-based data.

}

\maketitle
\color{black}

\section{Introduction}\label{sec:intro}
N-body numerical simulations \citep[e.g.][]{SP05.2} predict the
existence of dark matter halos, which contribute the knots within a
network of filamentary structure. This peculiar arrangement of matter
is frequently called the `cosmic web'. The appearance of these
filaments holds important information on the underlying cosmology and
structure formation history and constitutes an important tool to
investigate the main mechanisms of cosmic structure formation, along
with and confirmed by other observables \citep{1983MNRAS.204..891K,
  1987ApJ...313..505W,
  2005MNRAS.359..272C,2010MNRAS.408.2163A,2010MNRAS.406.1609B,2011MNRAS.416.3098F}.
The first observational evidence of such structures came from
spectroscopic surveys like the 2dF Galaxy Redshift Survey
\citep[see][and references therein]{2001MNRAS.328.1039C}, but these
data do not return direct information on the matter density
distribution. A mass-to-light ratio in filaments, which might be scale
dependent, must be assumed or estimated from independent
observations. Thus, it is important to find other means to evaluate
physical properties of filaments, especially when it comes to the
distribution and total amount of dark matter therein. The use of gravitational
lensing is, in principle, the most direct and elegant way to do so.
Cosmic shear surveys are already providing good constraints on
cosmological models by looking at the shear--shear correlation
function \citep[e.g.][]{kiblinger13, Schrabback2010}. Even if those
surveys are doing well in estimating the global properties of the
linear matter power spectrum, they are still failing in retrieving a
complete picture of the `cosmic web', although the path has been
outlined by the dark matter map of the COSMOS field \citep{Massey2007}
and recent results in the CFHTLenS \citep{vabwaerbeke13}. The
difficulties in getting direct measures from weak gravitational
lensing are due to the very weak deflection field which, so far, could
be measured statistically over very large portions of the sky but not
filament by filament in the field.  But exactly this information is most
important in the study of  the relation between galaxy clusters and
intra-cluster filaments.  However, promising progress has been made to
detect filaments between sub-clumps in merging clusters of
galaxies. Recent examples can be found in the super-cluster Abell
222/223 \citep{Dietrich2012} and the possible gas bridge connecting
pairs of clusters as detected with CMB observations \citep{PL12}.

Different weak lensing methods to detect filaments have been proposed
in the literature and  they are mostly based on a circular aperture
mass or on its second momenta
\citep{2005A&A...440..453D,2010MNRAS.401.2257M,2011A&A...532A..57S}.
These local lensing estimates are somewhat limited in their
sensitivity and so far have been applied to tight pairs of clusters
only, where the signal is expected to be higher, but at the cost of a
strong degeneracy with the signal induced by the main halos. In this paper we
overcome this limitation by providing a method which uses all
first-order lensing information coming from filaments as a whole and
which provides stronger signals, even with ground based data. We study
a specific case where a massive cluster is situated at the terminal
end of such a filamentary structure. A standard scenario, as suggested
by N-body numerical simulations. We show how our general approach is
favorable to control the strong cluster signal in a natural way and
able to highlight the structure of interest, the intra-cluster
filament and in general all matter anisotropies surrounding the
central cluster. Since this method is sensitive to the smooth
distribution of the dark matter only, we propose an additional
approach to exploit complementary information: The over-density of
dark matter clumps expected to align with the `cosmic web'. These two
different approaches should be used together since, on the one hand,
intra-cluster filaments could be seen as composed by a continuous
distribution of matter, and on the other hand as an ensemble of single
dark matter halos.

We structure our work as follows: In Section~\ref{sec:filamentLens} we
summarize the main properties of intra-cluster filaments relevant for
weak lensing, followed by Section~\ref{sec:intra-cluster} which
presents the weak lensing technique to infer their properties. In
Section~\ref{sec:sims} we test the method against numerical
simulations. The complementary approach to incorporate weak lensing
peak counts for filament detection is outlined in
Section~\ref{sec:wlcounts}. In Section~\ref{sec:conc} we draw our
conclusions.

\section{Lensing from intra-cluster filaments}\label{sec:filamentLens}

All observables of a cosmic structure related to gravitational lensing can be derived through the so
called lensing potential
\begin{equation}
  \psi(\vec{\theta}) \equiv \frac{2}{c^2}
     \frac{D_{\rm ds}}{D_{\rm d}D_{\rm s}}
     \int 
     \Phi(D_{\rm d}\vec{\theta}, z)\,dz ~.
\end{equation}
It characterizes the properties of intervening lenses between a given population of background sources and the observer. Here,
$\Phi$ defines the Newtonian potential of this matter distribution,
$\vec{\theta} = \{\theta_{1},\theta_{2}\}$ defines angular positions in the plane of the sky, $c$ is the speed of light,
and $D_{\rm ds}$, $D_{\rm d}$ and $D_{\rm s}$ are the lens--source,
observer--lens and observer--source angular diameter
distances, respectively. In this work we study only very weak
potentials, therefore we consider first order lensing quantities such
as the scalar convergence $\kappa$ and the complex shear $\gamma=\gamma_1+\mathrm{i}\gamma_2$,
which for any lensing potential $\psi$ are
\begin{eqnarray}
 2\gamma =&  \partial\partial \psi \;, \nonumber \\
 2\kappa =& \partial \partial^{*} \psi\;,
\end{eqnarray}
with the complex lensing operator $\partial :=
(\frac{\partial}{\partial\theta_{1}}
+\mathrm{i}\frac{\partial}{\partial\theta_{2}})$
\citep[see][]{2001PhR...340..291B, Schneider2008,Bartelmann2010}. The
main lensing observable is the complex reduced shear $g=\gamma/(1+\kappa)$,
but we can safely approximate $g$ with $\gamma$ since for filaments
the approximation $\kappa \ll 1$ holds with very high accuracy. In any
case, all equations presented in this study can be used with the
reduced shear instead of the shear whenever necessary.

\begin{figure}[!t]
  \centering
  \includegraphics[width=1.0\hsize]{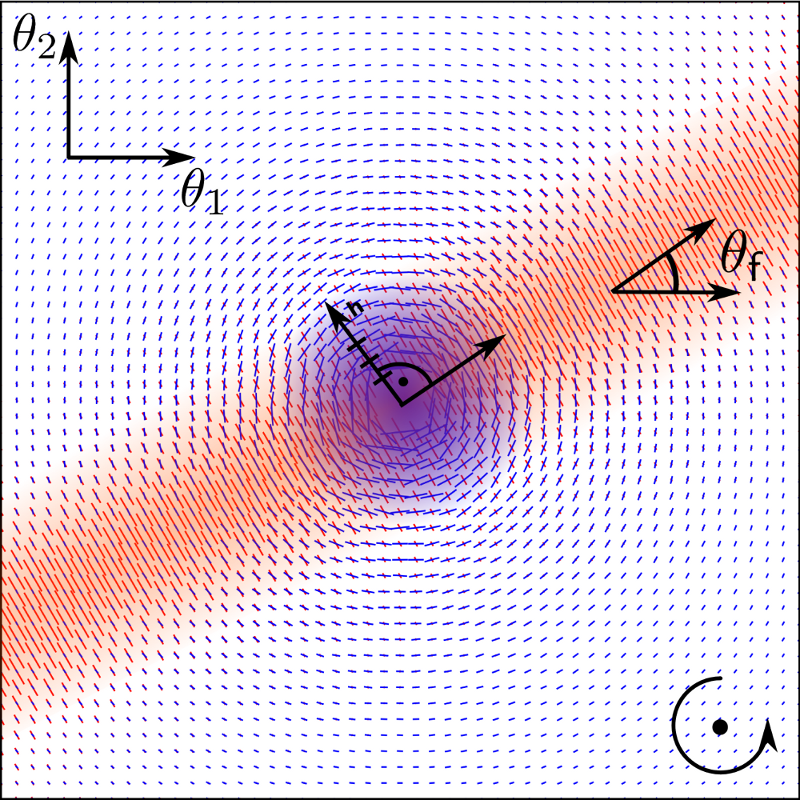}
  \caption{A schematic sketch of our Mock filament and cluster
    setup. The background whisker plot shows the shear fields of the
    filament (red) and the NFW halo (blue).  Note that the length of
    the filament whiskers is enhanced by a factor of 15 compared to
    the halo ones in order to make them more visible. The global
    coordinate system $\vec{\theta}$, the filament's orientation
    $\theta_{\textrm{f}}$ and the distance from the filament center
    along an axis perpendicular to its extension $h$ are also clearly
    marked. The 3D orientation of the coordinate frame is
    right-handed. In our analysis we treat the two sides of the
    filament as independent, so that we can access all possible
    geometrical configurations. The figure however shows the most simple configuration, being an identical orientation angle 
    for the two sides of the filament with respect to the main halo. }
  \label{fig:drawing}
\end{figure}

Once the general lensing quantities are defined, it is straightforward
to quantify the intra-cluster filaments by means of a simple mass
model which is shown in Figure~(\ref{fig:drawing}). For simplicity, we
assume the filaments to be of infinite length and that their matter
distribution is not varying along their major axis so that, for
symmetry reasons, the gravitational potential is constant along this
direction. Hence, the shear is aligned orthogonally to the filament,
i.e.  $\gamma_1=-\kappa$ and $\gamma_2=0$, assuming that the filament
falls on the respective Cartesian axis defined by the plane of the sky
$\vec{\theta}$. Generalizing for arbitrary filament directions
$\theta_{\rm f}$ and distances h to its major axis, we have
\begin{equation}\label{eq:gamma}
  \gamma=-\kappa(h) \left[\cos(2\theta_{\rm f})-i \sin(2\theta_{\rm f})\right]  \;.
\end{equation}

As mentioned before, we assume that the convergence is varying only
orthogonally to the filament direction and, following
\citet{2005MNRAS.359..272C} and \citet{2010MNRAS.401.2257M}, we assume
a convergence profile for the filament
\begin{equation}\label{eq:kappa}
  \kappa(h)=\frac{\kappa_{0}}{1+\left(\frac{h}{\mathrm{h_{\rm c}}}\right)^2} \;,
\end{equation}
where $\kappa_0$ is the filament's maximum convergence,
and $\mathrm{h_c}$ a scale
radius, representing its typical width (compare Figure~\ref{fig:drawing}).

\begin{figure*}[!t]
  \centering
  \includegraphics[width=0.48\hsize]{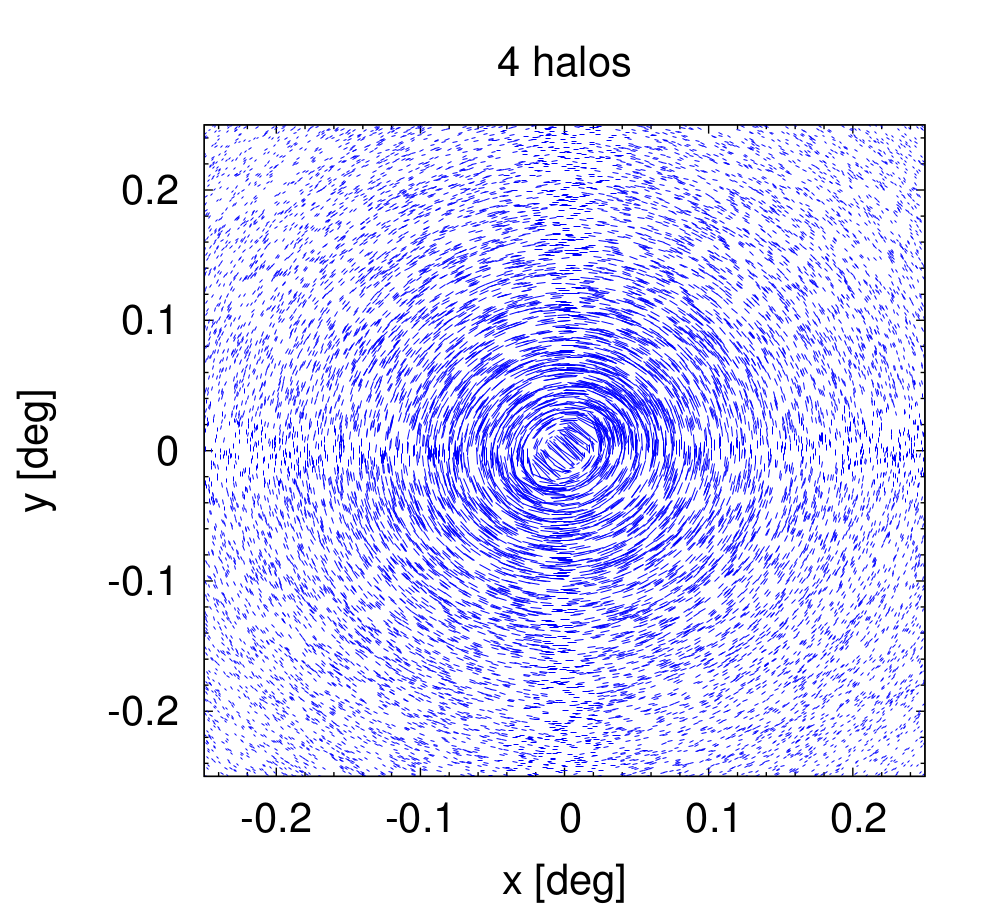}
  \includegraphics[width=0.48\hsize]{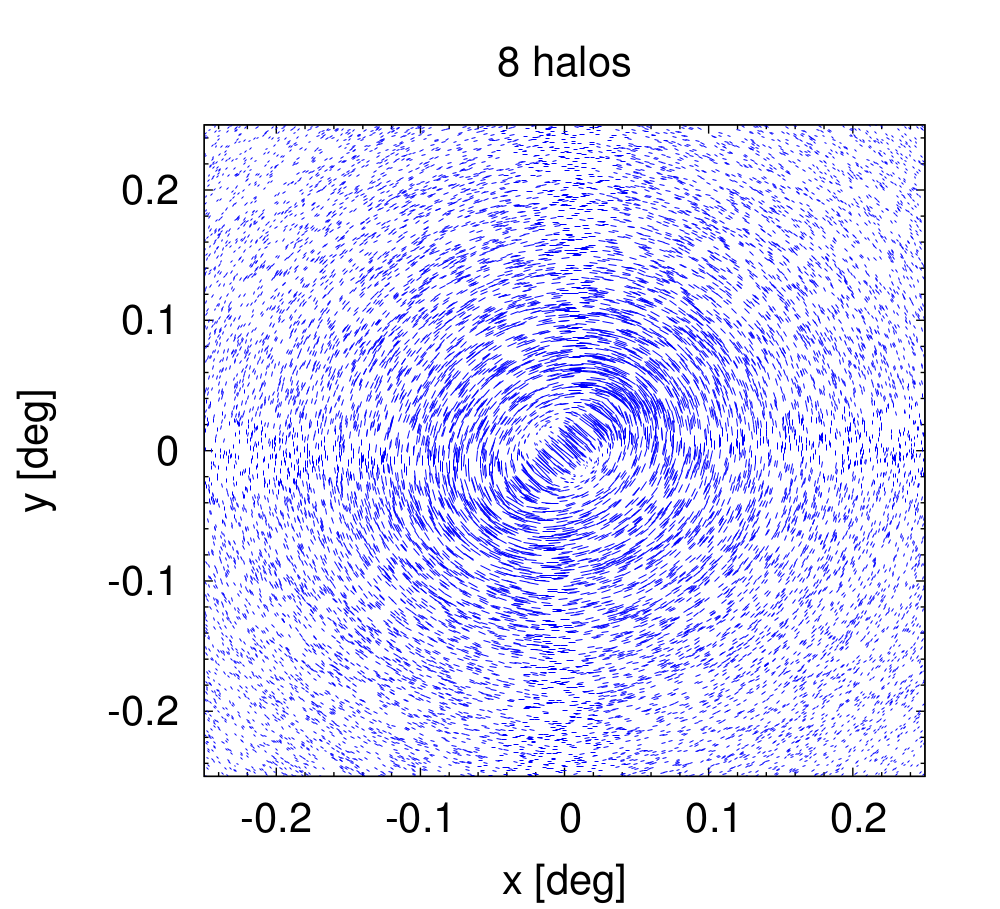}
  \includegraphics[width=0.48\hsize]{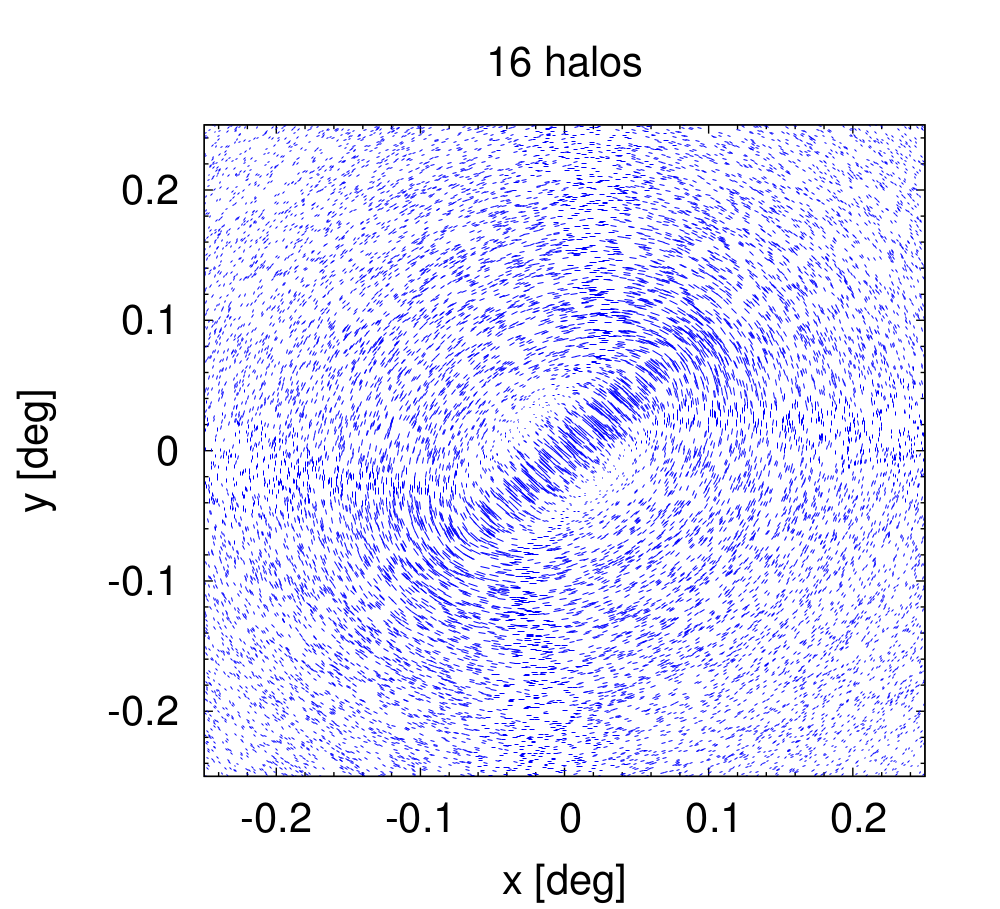}
  \includegraphics[width=0.48\hsize]{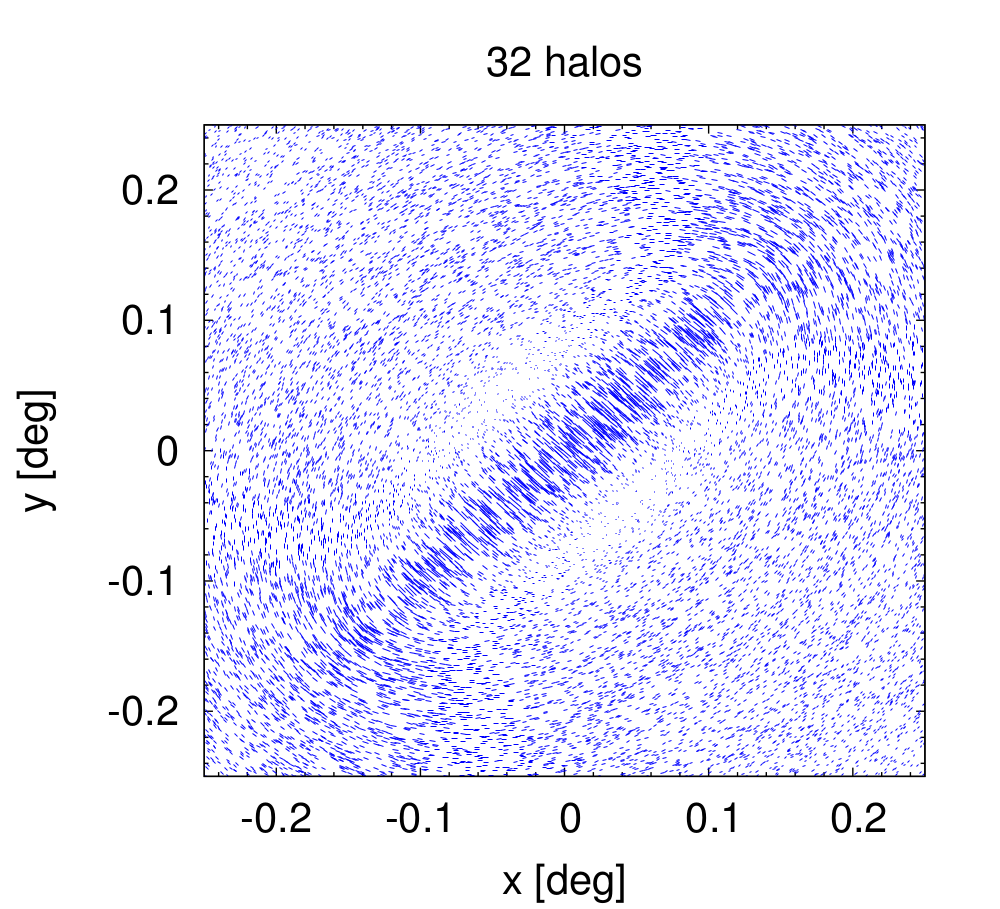}
  \caption{Shear pattern induced by an ensemble of 4, 8, 16 and 32 NFW
    halos with random masses ranging from 0 to $10^{15} M_{\odot}/h$.
    The halos are displaced along a line, forming an elongated
    structure resembling an intra-cluster filament located at
    $z=0.55$. The shear sampling is analogues to the one typical for
    weak lensing observations with a source number density of
    $n_s=15$~arcmin$^{-2}$. These toy models illustrate the two regimes
    in the shear pattern caused by elongated structures, i.e. an alignment
    'tangential' to the center of the
    structure at large radii and the alignment 'orthogonal' to the filament's major axis. The
    two regimes are separated by a boundary with vanishing shear.}
  \label{fig:NFW-filament}
\end{figure*}

To give a better understanding of this peculiar shear pattern, we plot
in Figure~(\ref{fig:NFW-filament}) four different structures located
at $z=0.55$ and composed by an ensemble of $4$, $8$, $16$ and $32$
halos, respectively. Their masses are randomly chosen, ranging from $0$ to $10^{15}
M_{\odot}/h$ and follow a Navarro--Frank--White (NFW) density profile
\citep{1997ApJ...490..493N,1996A&A...313..697B}.  All halos belonging to each structure 
are aligned along the respective field's diagonal
and are equally spaced by $0.5$ arcmin so that the object length grows
linearly with the number of halos. The shear is sampled such that it
resembles typical weak lensing observations with a source number
density of $n_s=15$~arcmin$^{-2}$. This specific representation of a
filament illustrates how the shear pattern caused by elongated
structures can be split into two regimes, one at large scales where
the shear is aligned tangentially with respect to the structure
barycenter, and one in the vicinity of the bar/filament where the
shear is displaced orthogonally with respect to the structure
elongation. A representative example of the latter case in the strong
lensing regime is given by the straight arc visible in the highly
elongated cluster ACT-CL J0102-4915 \citep{ZI13}, which finds its full
explanation in the structure layout we present here. Note how the
shear vanishes on the boundary between these two regimes. Clearly, the
approximations discussed in this section hold very well, justifying
their use in describing the main properties of intra-cluster
filaments, which are in return used for the actual filament detection.

\begin{figure*}[!t]
  \centering
  \includegraphics[width=0.455\hsize]{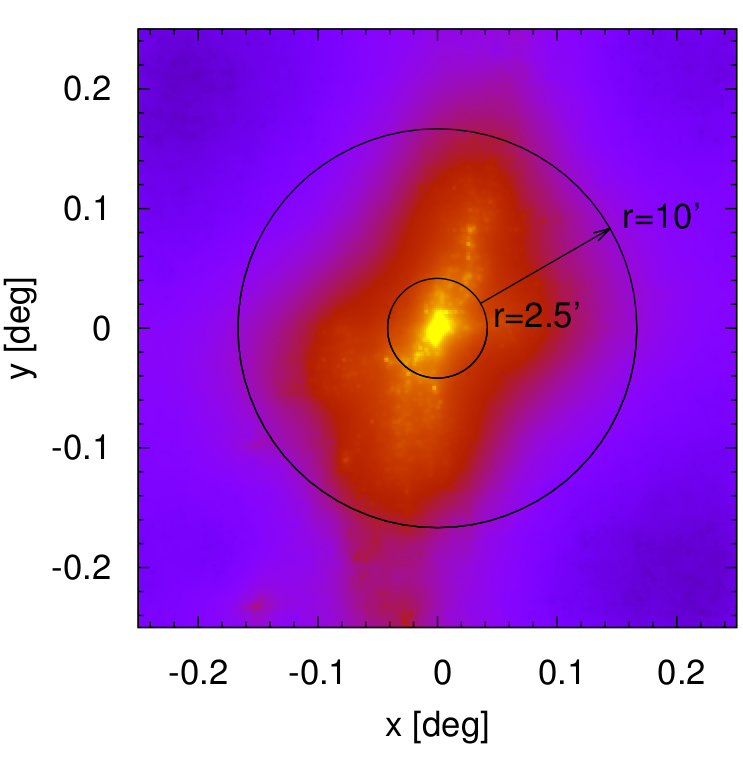}
  \includegraphics[width=0.457\hsize]{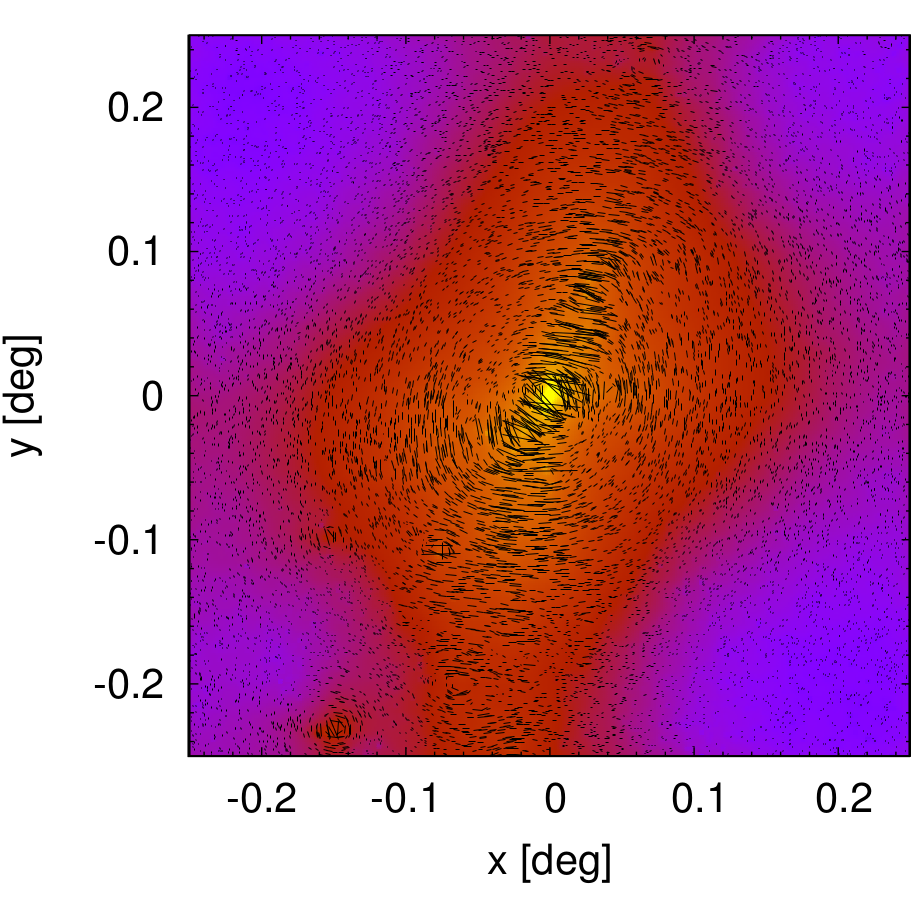}
  \caption{A numerical simulation of a cluster--filament configuration. Left panel: Shear amplitude map 
    centered on a massive galaxy cluster surrounded by
    filamentary structures.The circles enclose the area in which we
    measure the lensing signal for the filament detection. Right
    panel: Whisker visualization of the shear pattern of the same simulation. The shear of the
    central part of the main cluster was removed to preserve the
    clarity of the figure.}
  \label{fig:g696shear}
\end{figure*}

A realistic case of the expected signal is plotted in
Figure~(\ref{fig:g696shear}), which shows the convergence (left panel)
and shear pattern (right panel) of a N-body numerical simulation
carried out with GADGET-2 \citep{Springel2005,Springel2001},
containing a central massive, merging galaxy cluster together with a
extended filamentary structure. The simulation is based on a flat
$\Lambda CDM$ model with a present matter density parameter
$\Omega_\mathrm{m}=0.3$, a present baryon density parameter
$\Omega_\mathrm{b}=0.04$, a Hubble parameter $h=0.7$ and a power
spectrum normalization corresponding to $\sigma_8=0.9$. The number of
gas and DM particles, enclosed in the simulation box of ($50\times
50\times 70) ~\mathrm{Mpc}$, was $1.4\times 10^7$ and $1.7\times
10^7$, respectively. Their masses were set to
$m_{\mathrm{gas}}=2.43\times10^8 M_{\odot}$ and
$m_{\mathrm{DM}}=1.61\times 10^9 M_{\odot}$. The Plummer-equivalent
gravitational softening was set to
$\epsilon_{\mathrm{pl}}=7~\mathrm{kpc}$ comoving from $z=2$ to 0,
while it was taken to be fixed in physical units at higher redshifts
\citep[for more details on the simulation see][]{Dolag2006}. The shear
pattern known from Figure~(\ref{fig:NFW-filament}) is clearly visible
also in this realistic case, where four main filaments depart from the
central cluster. Note how in this case the filaments are aligned such
to create a `cross-like' structure with two non-orthogonal arms. Such
a configuration, with two connected filaments symmetrically extending
from two opposite sides of the central cluster is quite common and
constitutes 1/3rd of all cases of aligned filamentary structure in
cosmological simulations \citep{2005MNRAS.359..272C}. Nevertheless we
treat the two sides of the filament as independent, so that we can
access all possible geometrical configurations.

\section{Intra-cluster filament detection}\label{sec:intra-cluster}

The very small shear signal expected for filaments \citep[e.g.][]{Dolag2006}
clearly states the challenge in detecting and quantifying the dark
matter properties of such structures. This is mainly due to
observational noise, large scale structure (LSS) contamination
(defined as the population of all filaments along the line-of-sight
except the one under investigation) and the massive galaxy clusters
expected to sit along their major axes. To overcome these
difficulties, we try to fully exploit the peculiar morphological
properties of the setup and the large area covered by
filaments. Hence, we aim at using global lensing properties as data input
instead of measuring local quantities.

Specific filament configurations with respect to the surrounding
galaxy clusters have to be identified and considered since, if
favorable, they might provide the crucial piece of information to help
the detection and to quantify the dark matter content.  Therefore, we
restrict our search on a specific configuration which is still general
enough to be applied to the majority of the observable cases.  We
focus our attention on the case in which a large galaxy cluster is
sitting at the terminal end of a straight filament.  This is a very
general case \citep{2005MNRAS.359..272C}, where the cluster helps to
define a physical coordinate frame and has symmetry properties, as we will show later on,  
greatly enhancing the separation of the cluster and filament components.

\subsection{A shear decomposition for filaments}\label{sec:decompose}

The shear signal of intra-cluster filaments is aligned with their
major axis as discussed in Section~(\ref{sec:filamentLens}) and shown
in Figures~(\ref{fig:drawing}), (\ref{fig:NFW-filament}) and
(\ref{fig:g696shear}). It is thus convenient to decompose the shear in
a tangential and a cross component relative to the
filament direction
\begin{equation}\label{eq:decompose}
  \gamma_{\rm q} =
  -\left[
  \gamma_1\cos(2\alpha_{\rm q}) + \gamma_2\sin(2\alpha_{\rm q})
  \right] \;,
\end{equation}
where $\gamma_{\rm q}$ stands for the tangential or cross shear
components with $\alpha_{\rm t}=\theta_{\rm f}-\pi/2$ and
$\alpha_{\rm x}=\theta_{\rm f}-\pi/4$, respectively (see
Figure~\ref{fig:drawing} for visual orientation). If we now use the
filament model defined by Equations~(\ref{eq:gamma}) and
(\ref{eq:kappa}) we obtain
\begin{equation}
  \gamma_{\rm f,t}(\vec{\theta}) = \kappa(h) \;\;\mbox{,}\;\;
  \gamma_{\rm f,x}(\vec{\theta})=0
\end{equation}
where, in radial coordinates, $h=r\cos(\theta-\theta_{\rm f})$. The same
decomposition has to be applied to all contributing matter components including the
central cluster, for which
\begin{equation}
  \gamma_{\rm c,t}(r,\theta) = g(r)\cos(2(\theta-\theta_{\rm f})) \;,
\end{equation}
\begin{equation}
  \gamma_{\rm c,x}(r,\theta) = g(r)\sin(2(\theta-\theta_{\rm f})) \;,
\end{equation}
and where $g(r)$ is the cluster's reduced shear radial profile, which we
assume to follow a NFW profile. Note that, in
contrast to the filament itself, for the central part of the cluster the
approximation $\gamma\ll 1$ is not valid and the reduced shear has to
be used.

In order to perform this decomposition, the filament direction
$\theta_{\rm f}$ must be known in advance.  We will explain later how
the input lensing data can be directly used to infer this
information. For the moment and simplicity of the argument, we assume
to have perfect knowledge of $\theta_{\rm f}$.

\subsection{Optimal filter}\label{sec:optimal}

From now on, we use the decomposition scheme described in
Section~(\ref{sec:decompose}) together with a filtering approach
directly applicable to input weak lensing catalogs containing galaxy ellipticities. We follow
the recipe proposed by \cite{MAT04.2}, which is based on physically
motivated quantities such as the expected filament shear profile, the
cosmic shear power spectrum and the survey properties. This method
avoids the use of polynomial or Gaussian weight functions for which
the parameters have to be determined empirically with the help of expensive
numerical simulations and which generally do not ensure optimized results
with respect to the signal-to-noise ratio
\citep{MAT05.3,2007A&A...471..731P}. It is a linear matched filter,
usually applied under the assumption of circular symmetry and in close analogy to the aperture mass
\citep{Fahlman1994,1996MNRAS.283..837S}. However, in this work, we consider
the filament geometry which is far from being circular.
In addition to this optimal filter, we show how the adopted
decomposition provides a natural way to control the central cluster
contribution.

\begin{figure}[!t]
  \centering

  \includegraphics[width=0.98\hsize]{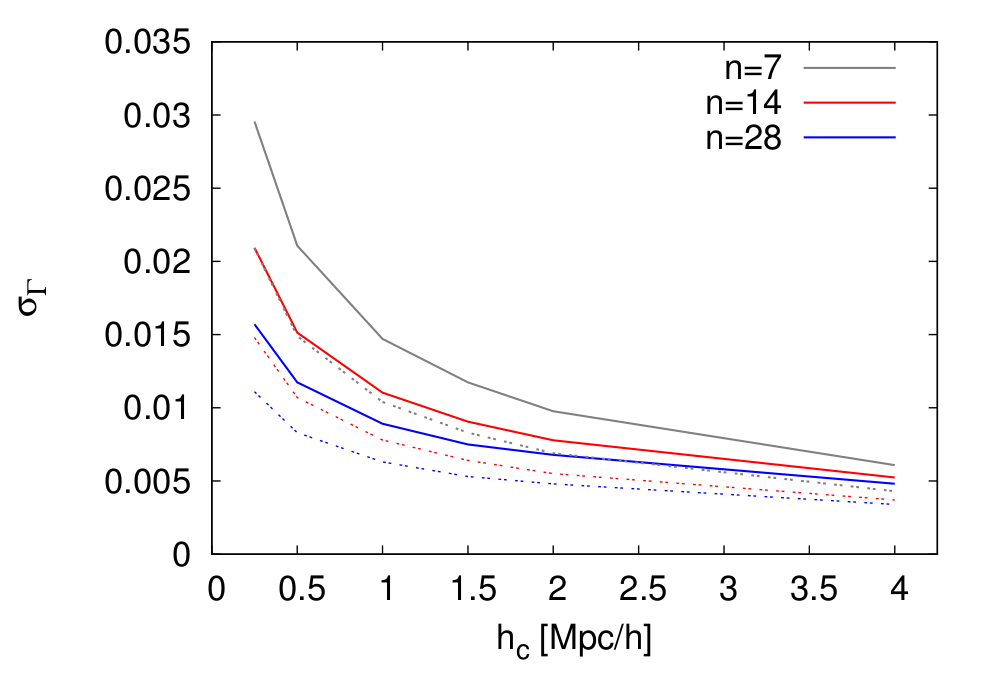}
  \includegraphics[width=0.98\hsize]{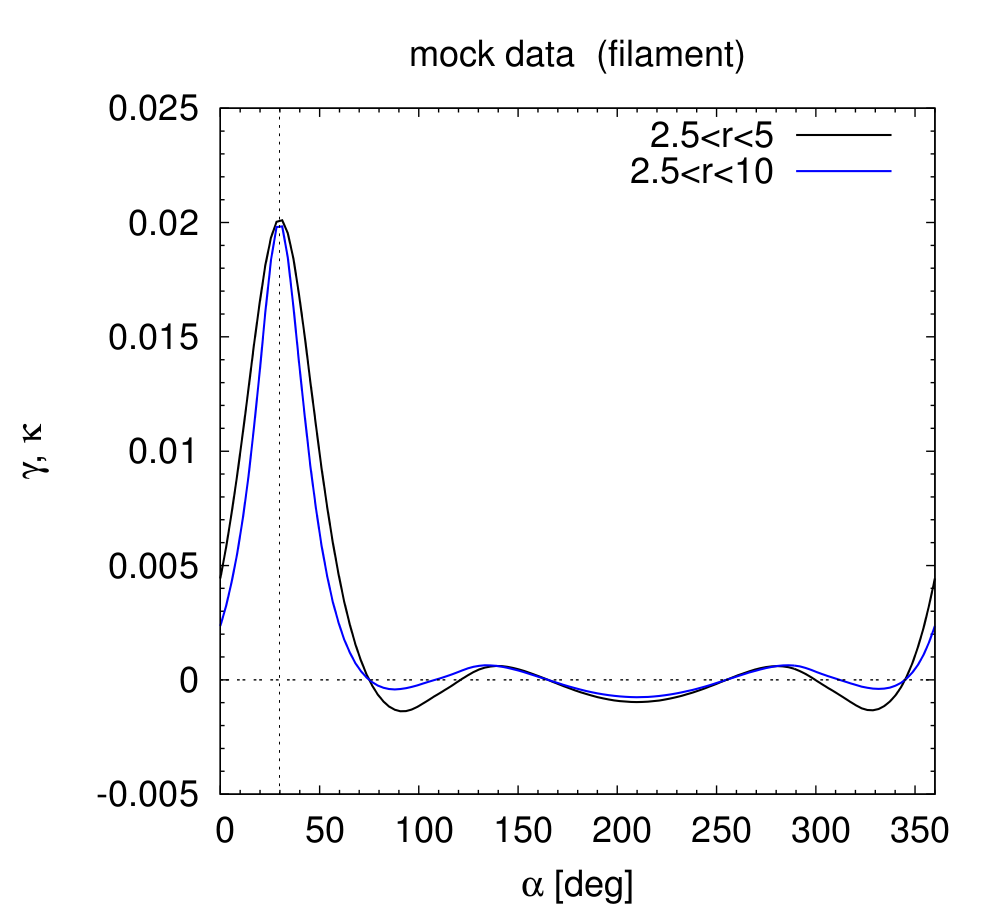}
  \caption{An optimal linear filter for filament detection. Top panel: Variance for filaments at redshift $z=0.55$. For
    a given survey with intrinsic ellipticity scatter of $\sigma_{\rm \epsilon}=0.4$, different core radii
    $h_c$ and different galaxies number densities $n$ are
    plotted. Continuous lines refer to a filter with length $L=10$
    arcmin while dashed lines refer to $L=20$ arcmin. Bottom panel: Expected
    filter response in the presence of a filament located at $z=0.55$,
    with a major axis position angle of $\theta=30$ degrees, a maximum
    convergence of $\kappa_0=0.02$ and a scale radius of $h_c=0.25$
    Mpc/h. The filter was truncated with an inner cut-off of $2.5$
    arcmin and two different outer cut-offs of $5$ and $10$ arcmin.  }
  \label{fig:sigma}
\end{figure}

To obtain the filter function, we base its derivation on a data model,
$d$, tailored towards our case
\begin{equation}\label{eq:signal}
  d_{\rm q}(\vec{x}) = \gamma_{\rm c,q}(\vec{x}) + \gamma_{\rm
    f,q}(\vec{x}) + n_{\rm q}(\vec{x}) \;,
\end{equation}
where $\gamma_{\rm c,q}$ and $\gamma_{\rm f,q}$ are the cluster
(isotropic) and filament (anisotropic) signals already discussed and
$n$ is the random noise contribution given by the LSS contamination,
shot noise and intrinsic ellipticity of the galaxies. Again, $\rm q$
stands for the tangential or cross shear components, $\rm t$ and $\rm
x$, respectively. The noise component, $n$, can be modeled as an
isotropic Gaussian random field with zero mean so that it is fully
described by its power spectrum $P(k)$:
$\left<n(\vec{k})n(\vec{k}^\prime)\right>=(2\pi)^2\delta(k-k^\prime)P(k)$. According
to Equation~(\ref{eq:kappa}) the filament signal is linear in
amplitude, $\kappa_{0}$. Therefore, it is convenient to define a
linear estimator which in most general form is given by
\begin{equation}\label{eq:estimator}
  \Gamma_{\rm q}(\vec{x})=\int
  d_{\rm q}(\vec{x}^{\prime})\Psi(\vec{x}-\vec{x}^{\prime})\d^2x^{\prime} \;.
\end{equation}
with an outcome variance
\begin{equation}\label{eq:variance}
  \sigma^2_\Gamma = \int \frac{\d^2k}{(2\pi)^2} P(k)\Psi^2(\vec{k}) \;,
\end{equation}
conveniently derived in the Fourier space.

The optimal filter function, $\Psi$, is derived through a constrained
minimization where the variance of the measure has to be minimal, i.e.
$\partial\sigma_{\Gamma}^2/\partial\Psi=0$, while returning unbiased
results with respect to $\kappa_{0}$, i.e. $\left<\kappa_{0}\right>=\Gamma_{\rm
  t}(\vec{0})$ and reads
\begin{equation}\label{eq:filter}
  \hat \Psi(\vec k) =
  \alpha
  \frac{\hat\tau(\vec{k})}{P(k)}
  \quad\mbox{with}\quad
  \alpha^{-1}=
  \int \d^2 k \frac{\left|\hat\tau (\vec{k})\right|^2}{P(k)}\;,
\end{equation}
where $\hat\tau(k)$ is the Fourier transform of the expected filament
tangential shear, expressed as $\gamma_{\rm
  f,t}(\vec{x})=\kappa_{0}\tau(\vec{x})$. $P(k)=P_{\rm g}(k)+P_{\rm
  LSS}(k)$ is the noise power spectrum including the linearly evolved
LSS through $P_{\rm LSS}$,
as well as the source shot noise and intrinsic ellipticity $P_{\rm
  g}(k)=\sigma_{\rm \epsilon}^2/(2n_{\rm g})\exp\left(k^2/(n_{\rm
  g}\ln2)\right)$, given their angular number density $n_{\rm g}$ and
the intrinsic ellipticity dispersion $\sigma_{\rm \epsilon}$
\citep{MAT04.2}. The exponential in the $P_{\rm g}$ term accounts for
the average background source separation which limits the angular resolution
achievable by the filter and in general by any method.

In the top panel of Figure~(\ref{fig:sigma}) we evaluate the noise
variance for different filament widths, lengths and general survey
properties. In this example we consider a structure located at
$z=0.55$ and for the noise contribution expressed by the LSS power
spectrum we adopt a $\Lambda$CDM model with a matter over-density, a
cosmological constant and an Hubble expansion rate of $\Omega_m=0.3$,
$\Lambda=0.7$ and $h=0.7$, respectively. We further assume the sources
to be placed at redshift $z=1.0$, we fix their intrinsic ellipticity
to $\sigma_{\rm \epsilon}=0.4$ and we truncate the filter length to 20
arcmin, i.e. $\sim 5.5$ Mpc/h at this redshift, a common value for
current wide field ground based imagers, to avoid possible misleading
features given by a square field of view. The variance decreases for
larger template width $\mathrm{h_{\rm c}}$ because more sources are
involved in the estimate reducing the shot noise and intrinsic
ellipticity scatter. Note that the variance is not changing with the
square root of $n$ because of the LSS component, which is constant for
any survey.

\subsection{A natural way to control the central cluster contribution}\label{sec:naturally}
\begin{figure*}[]
  \centering
  \includegraphics[width=0.48\hsize]{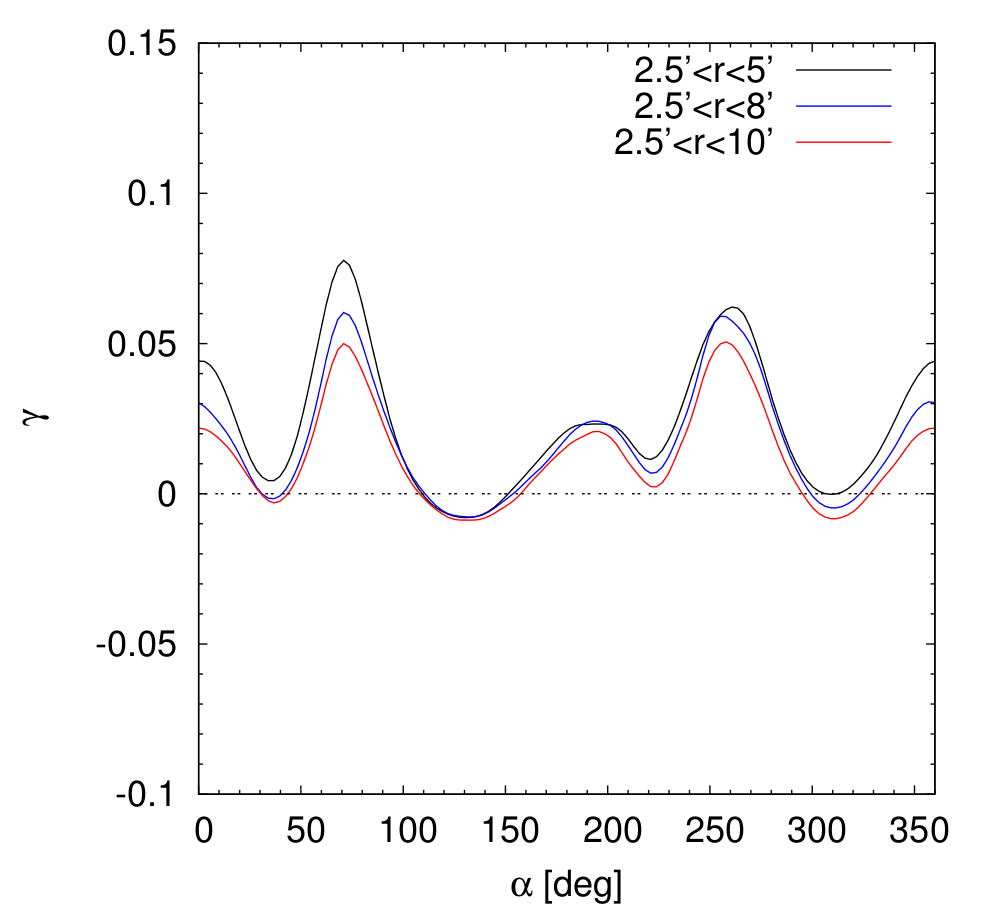}
  \includegraphics[width=0.48\hsize]{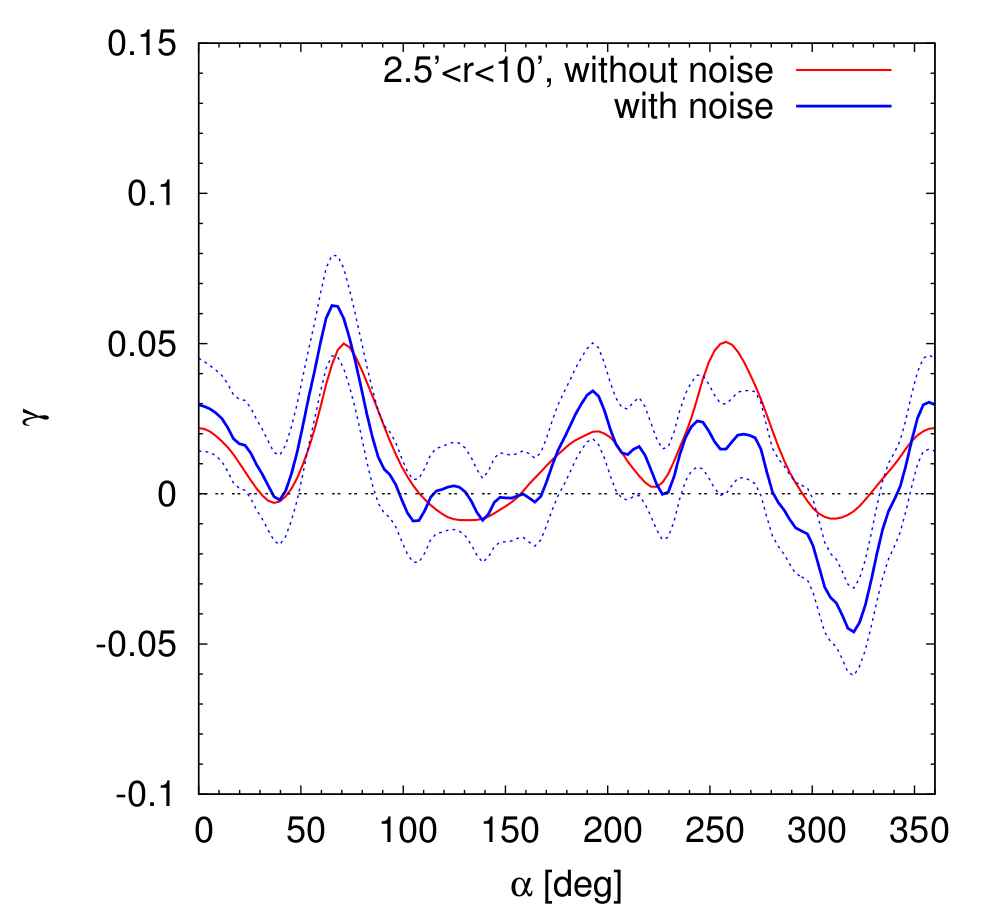}
  \caption{The optimal linear filter applied to realistic data. Left panel: The filter response on a mock
    catalog based on the N-body numerical simulation without noise and 
    as a function of scanning angle. Each line represents the result
    for three different radial cut-offs. Right panel: This case
    considers the noise given by a finite source number density of
    $n_s=15$~arcmin$^{-2}$ and an intrinsic ellipticity r.m.s. of
    $\sigma_g=0.4$. The red line represents the expected theoretical
    signal (the same as in the left panel), the blue line the filter
    outcome for noisy data and the dashed blue lines its $\sim1\sigma$ upper and lower
    uncertainty. The most significant filamentary structure (at
    $\alpha\sim 70$ deg) has been recovered at the $\sim4\sigma$ level, all
    others are detected with a lower significance.}
  \label{fig:numsim-filter-result}
\end{figure*}

The applied filtering scheme is explicitly designed to maximize the
filament's contribution to the signal-to-noise ratio with respect to the stochastic noise,
but it can also take into account, even if not explicitly, the central
cluster contamination which is in our case another source of noise.
On the one hand, by taking advantage of the filter's linearity, we
plug Equation~(\ref{eq:signal}) into Equation~(\ref{eq:estimator}) so
that we are left with the following three distinct components
\begin{equation}
  \Gamma_{\rm q}=\Gamma_{\rm c,q}+\Gamma_{\rm f,q}+\Gamma_{\rm n,q} \;,
\end{equation}
where $\Gamma_{\rm n,q}$ is a Gaussian random quantity, characterized
by Equation~(\ref{eq:variance}). On the other hand, the shear
decomposition scheme described in Equation~(\ref{eq:decompose}) allows
to separate the anisotropic signal component, given by the
filament, from the isotropic one associated to the central
cluster. In fact, the filter is asymmetric due to the highly elongated
template $\tau$, describing the filament shape and therefore,
  the outcome of applying the filter to a given filament configuration
  strongly depends on its orientation $\theta_{\rm f}$. Hence, it is necessary 
  to scan the filter through all possible orientation angles.

As a natural choice, the center of the reference frame is place on a massive galaxy
cluster because it marks the center around which we expect to find
filamentary structures departing from its center.  Once the filter is
centered in the reference frame, we scan all angles in a standard
right-handed Cartesian system; see Figure~(\ref{fig:drawing}). By
doing so, the measured tangential components of the filament as a
function of the spanned angle $\theta$ has a maximum, $\Gamma_{\rm
  f,t}(\theta=\theta_{\rm f}) = \kappa_{0}$,
while the cross component is zero, $\Gamma_{\rm f,x}(\theta=\theta_{\rm f}) =
0$. This is shown in the bottom panel of Figure~(\ref{fig:sigma}),
where we plot the expected filter response in the presence of a
filament located at $z=0.55$, with a major axis position angle of
$\theta=30$ degrees, a maximum convergence of $\kappa_0=0.02$ and a
scale radius of $h_c=0.25$ Mpc/h \citep[typical values derived from
  numerical simulations,
  see][]{2005MNRAS.359..272C,2010MNRAS.401.2257M}. This signature
indicates the filament orientation and a direct estimate of the
filament convergence $\kappa_{0}$. This behavior is a distinctive
indicator to discriminate the filament's signal components in contrast
to the contribution from the main central halo, $\Gamma_{\rm
  c,t}(\theta) \approx $~const. and $\Gamma_{\rm c,x}(\theta) \approx 0$, which
are nearly constant (nearly because clusters are just approximately circularly symmetric) for
all angles $\theta$.  With this approach we maximize the filament's
signal-to-noise ratio with respect to the survey properties by using
the \cite{MAT04.2} optimal filter and control the cluster signal,
which turns out to be nearly constant for any angle $\theta$.  A
spurious oscillatory behavior resulting from the cluster could be due
to an eventual presence of cluster ellipticity but, if this is the case,
it is always possible to mask a circularly symmetric area centered on
the cluster to reduce its significance. An alternative, and perhaps more conservative, way 
to approach the interpretation of the outcome signal is to
simply consider its isotropic and anisotropic components regardless
their association to the filament or the central cluster, given the
fact that these two are physically tightly related, actually
constituting a single system.

Summarizing, the strategy leads to the final, simple check-list for
filament detection:
\begin{enumerate}
  \item Define the reference coordinates with respect to a massive
    galaxy cluster,
  \item Apply the optimal filter described in
    Equation~(\ref{eq:filter}) to the tailored shear
    components of the input data (Equation~\ref{eq:decompose}) and sample all position
    angles,
  \item All anisotropic signal components will be picked up as
    oscillations and maxima. If a straight, intra-cluster filament is
    detectable, a relatively sharp maximum is expected in the
    tangential component, $\Gamma_{\rm t}$, in coincidence with the
    filament major axis direction.
\end{enumerate}

%\section{Filament detection with an optimal filter} \label{sec:sims}

\section{Testing the filter on a realistic N-Body numerical simulation} \label{sec:sims}

To test our novel filtering approach, we realized a Mock catalog of
galaxies lensed by the halos and filamentary structures present in
the N-body numerical simulation described in
Section~(\ref{sec:filamentLens}). The contribution of LSS was
neglected for the source catalog construction but its power spectrum
was used in the filter definition. Throughout the following analysis, we present a
case study for a system at redshift $z=0.55$ and an intrinsic
ellipticity dispersion of $\sigma_{\rm \epsilon}= 0.4$. The chosen
source number density of $n=15~\textrm{arcmin}^{-2}$ and a
field of view of $(50\times 50)$ arcmin are typical values for ground based
observations of a cluster field. In this case, we apply an external
circular truncation to the filament template with an exponential drop
to avoid artifacts given by the square field of view. The filter was
renormalized accordingly to preserve its unbiased properties presented
in the previous sections.

Since we are dealing with discrete sources, it is necessary to
approximate Equations~(\ref{eq:estimator}) and (\ref{eq:variance}) by
summing over the galaxy ellipticities, which are an estimator of the
shear
\begin{equation}
  \Gamma_{\rm q}(\theta) = \frac{1}{n}\sum_i \epsilon_{\rm
    iq}\Psi(\vec{x};\theta) \;,
\end{equation}
\begin{equation}
  \sigma^2_{\rm \Gamma_q}(\theta) = \frac{1}{2n^2}\sum_i |\epsilon_{\rm
    iq}|^2\Psi^2(\vec{x};\theta) \;.
\end{equation}
Here $\epsilon_{\rm iq}$ denotes the tangential or cross component of
the $i$-th galaxy ellipticity relative to the shortest line connecting
its location to the filament major axis, see
Equation~(\ref{eq:decompose}).

\begin{figure}[]
  \centering
  \includegraphics[width=0.98\hsize]{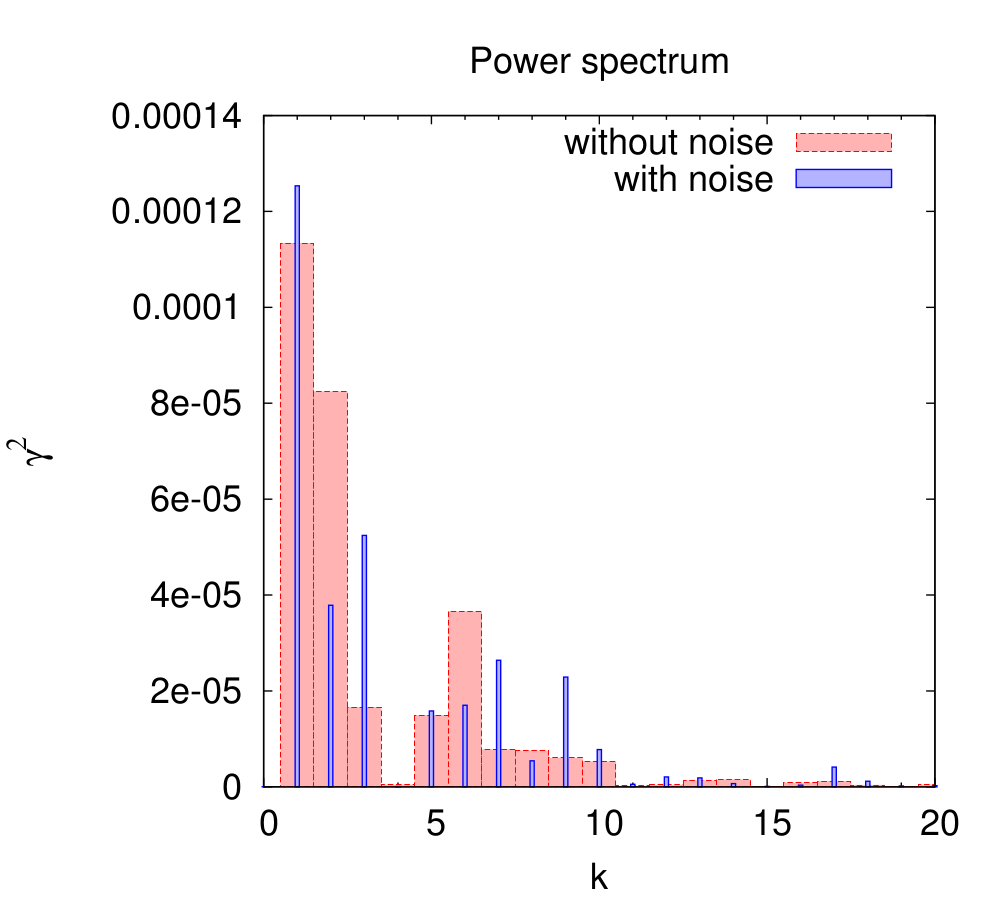}
  \caption{Signal power resulting from a Fourier analysis of the
  filter applied to the N-body numerical simulation. Red bars represent results in the absence of 
  intrinsic ellipticity noise of galaxies, blue bars take it into account.}
  \label{fig:NbodyResultsPeriod}
\end{figure}

The results of the filter applied to this mock simulation with and
without the intrinsic ellipticity noise of the background galaxies and
for different radius cut-offs are shown in the left and right panels
of Figure~(\ref{fig:numsim-filter-result}), respectively.  We
implemented a filter template with a scale radius of $h_{\rm c}=0.25$
Mpc/h, a total length of $L=10$ arcmin and an inner truncation of
$2.5$ arcmin to avoid the central cluster contamination. These radial
cut-offs are shown in Figure~(\ref{fig:g696shear}). The expected
behavior, i.e. a signal peak in conform with the position angles (PA)
of the filamentary structures around the central cluster at
$\theta\approx70$ and $\theta\approx260$ degrees, is seen with a
significance of $\sim 4 \sigma$ and $\sim1\sigma$, respectively. Also
the two smaller elongated structures at $\theta\approx0$ and
$\theta\approx190$ degrees are picked up by the filter at a $2\sigma$
level, showing its sensitivity. We note that the expected theoretical
r.m.s including the LSS contamination for this case is
$\sigma_\Gamma=0.015$ as derived by Equation~(\ref{eq:variance}).  In
this realistic case, also the ellipticity of the central cluster plays
a role in the signal since the filter is sensitive to all anisotropic
contributions. Even if the cluster elongation is physically correlated
to the presence of a filament, both in direction and amplitude as seen
in N-body numerical simulations, we tried to minimize its contribution
by applying a circular mask of $2.5$ arcmin in radius centered on the
main cluster. We do this, because we are aiming at the filament alone
even if its tight connection to the central cluster would be
physically motivated and in fact the study of the signal anisotropy
around clusters would be a sensible question to answer with this
method.  Another representation of this result is given in
Figure~(\ref{fig:NbodyResultsPeriod}), where we performed a Fourier
analysis of the filter outcome to highlight the most important scales
involved.

  Even if a single, ``massive'' filamentary  structure can be detected at a $4\sigma$
  level, the characterization of an
  anisotropic component around clusters can be carried out in a
  statistical sense and over a large number of clusters. One obvious way is 
  stacking, but this would
  lead to biased results towards higher filter responses. 
  A more promising
  approach is to count the maxima of the signal response instead,
  in analogy to the shear peak count statistics, allowing an unbiased
  estimate which is representative of the filament population in the vicinity
  of clusters. Moreover this would allow to separate the actual
  signal from the contribution of matter distributed along the line
  of sight which is uncorrelated to the investigated
  structures.

%\section{Filament detection with weak lensing peak number counts}\label{sec:wlcounts}

\section{An alternative method based on peak counts}\label{sec:wlcounts}

We have shown that a relatively continuous and smooth distribution of
matter, organized in a filamentary structure, can be detected down to
$\kappa_0\sim0.01-0.005$. This number assumes a source number density of
$n=15~\textrm{arcmin}^{-2}$ and an intrinsic background source
ellipticity dispersion of $\sigma_{\rm \epsilon}=0.4$, which are
typical values for ground based observations. Although the suggested
method seems to be sensitive to intra--cluster filaments even with
ground based observations, we suggest an additional, complementary approach to
search for filamentary structures. This different technique is based
on shear peak counts \citep{MAT09.1,DI10.1,MA11.1,marian13} and is
motivated by N-body numerical simulations, which predict an overdensity
of collapsed substructures along these filaments which could be
directly correlated with other observables such as optical cluster
detection methods \citep[see e.g.][]{2011MNRAS.413.1145B}.  Using
this method, together with the proposed filtering approach, poses the
interesting task of comparing the ratio of the dark matter distributed
in the smooth filament component, composed by an ensemble of numerous
halos of galaxy/group size, and in the distinct clumps, of more
massive halos, along its major axes.

\begin{figure*}[!t]
  \centering
  \includegraphics[width=0.48\hsize]{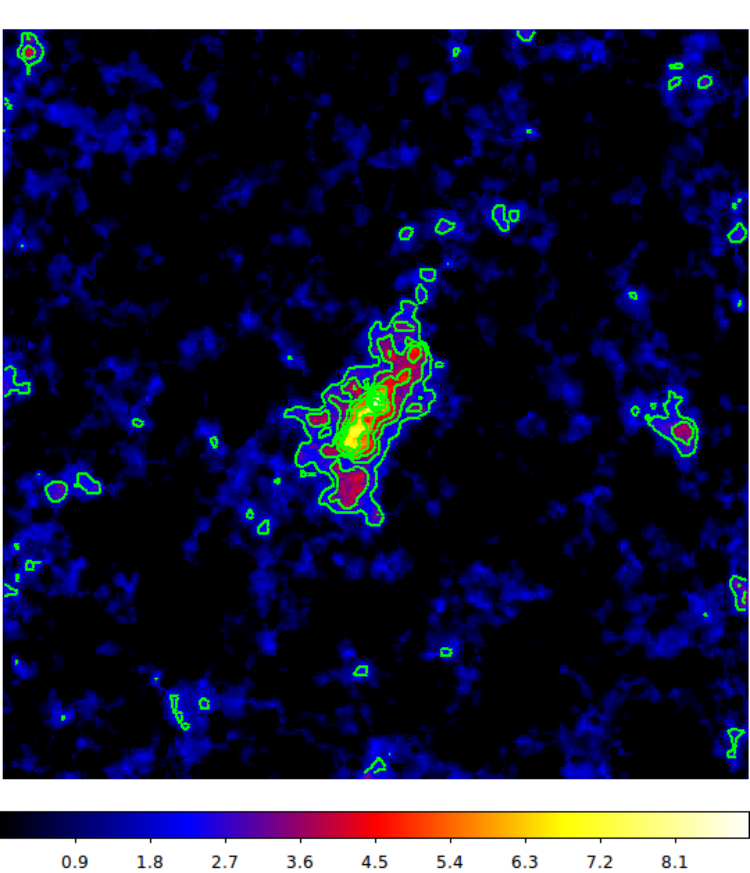}
  \includegraphics[width=0.48\hsize]{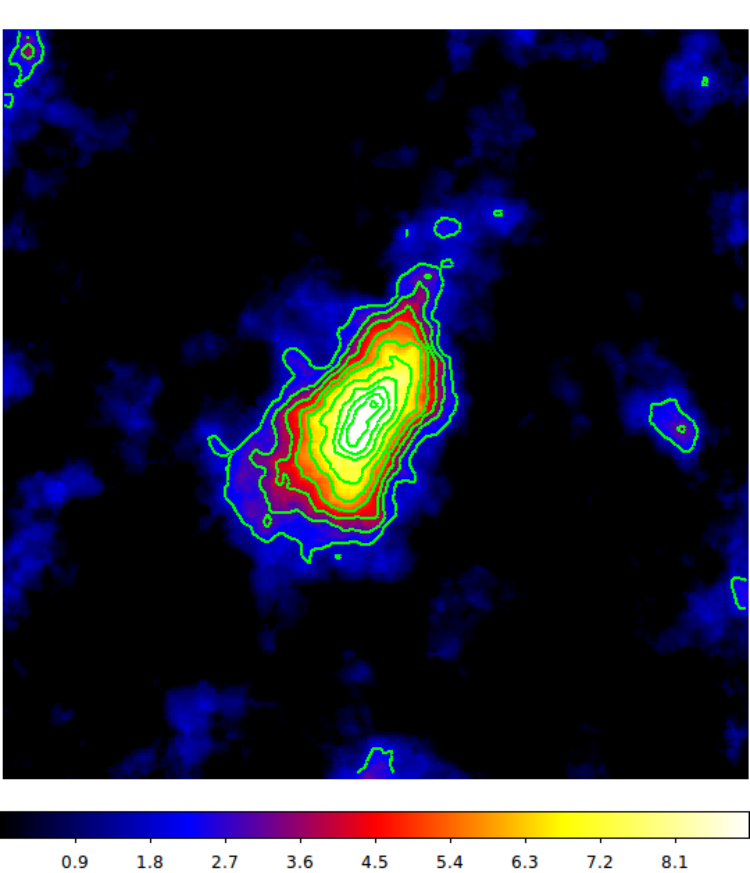}
  \caption{Weak lensing analysis of the N-body numerical simulation \citep{Dolag2006}
    performed with the isotropic, optimal matched filter described in
    \cite{MAT04.2}. The filter was optimized to detect NFW dark matter
    halos with masses of $5\times10^{13}M_{\odot}/h$ (left panel) and
    $1\times10^{15}M_{\odot}/h$ (right panel). The shown isocontours
    start from $S/N=2$ in steps of $\Delta_{S/N}=1$. The main halo
     and its ellipticity is clearly visible in the center of the field, 
  but also the additional small sub-structure in the field.}
  \label{fig:NBodyMap}
\end{figure*}

To detect collapsed structures, again we apply the recipe described by
\cite{MAT04.2} but in its original form, which is an aperture mass
filter designed to maximize the signal-to-noise ratio of the lensing
signal arising from dark matter halos. It takes into account the
general survey properties and the presence of LSS.  We tuned this
filter to search for NFW (sub-) structures with a mass of
$m=5\times10^{13}M_{\odot}/\textrm{h}$ and which are located at the main
cluster's redshift.  Different masses and redshifts could be used but
we focused our search on relatively small halos which are more likely
to belong to filaments. If smaller masses, i.e. scales, are used as
input template for the filter, its optimization criteria will prevent
it to act on excessively small angular scales. This is to keep the
source shot noise and intrinsic ellipticity minimal.

We apply this isotropic filter to the Mock catalog derived from the previous
N-body numerical simulation. We show the signal-to-noise ratio map in
Figure~(\ref{fig:NBodyMap}), together with the results for the same
filter but tuned to detect halos with $M=10^{15}\;M_{\odot}$, for
comparison. The isocontours represent signal-to-noise ratios larger
than $S/N=2$ in steps of $\Delta_{S/N}=1$. The most significant
structures are aligned along an axis centered on the main cluster and
with an inclination of $\sim60~\textrm{deg}$, showing a clear
correlation of collapsed structures with the filament. Other minor
structures are also visible in the field of view. To quantify the
significance of this correlation, we evaluated the number density of
detections comprised in a rectangular area centered on the larger
filament detected in Section~(\ref{sec:sims}), i.e. the one pointing
in the $\theta=60$ degrees direction.  This
rectangular area excluded the central cluster within a distance of
$r=2.5$ arcmin and its width was set to $7$ arcmin, a typical value
found in N-body numerical simulations \citep[see e.g.][for
  details]{2010MNRAS.408.2163A,2010MNRAS.406.1609B,2011MNRAS.416.3098F,2005MNRAS.359..272C}. The
other smaller structures are not detectable with other commonly used
weak lensing techniques.  The resulting overdensity contrast with
respect to the rest of the field, defined as the remaining area, is
$f=2.66\pm0.63$.

\section{Conclusions}\label{sec:conc}

We investigate the possibility to detect straight filaments centered
on a single massive cluster with actual ground-based weak lensing
data. In this lensing approach we solely focus on the dark matter
content and on filaments which depart radially from a massive
cluster. We choose this specific filament/cluster configuration
because it constitutes the large majority of cases \citep{
  2005MNRAS.359..272C} and because it has convenient geometrical
properties which help their detection through weak lensing. We
demonstrated this last aspect by analyzing a Mock catalog of galaxy
ellipticities based on a N-body numerical simulation containing a
massive galaxy cluster with connecting filamentary structures.

For this purpose, and in particular for the detection of the smooth
component of the filament's matter distribution, we adapted the
\cite{MAT04.2} optimal matched filter and modified it to serve our
specific case. We use a template tailored to the elongated signal of
straight filaments, together with a peculiar shear decomposition which
splits the shear in two components, one orthogonal and one
perpendicular to a given direction chosen such to verify the presence
of a filament in the field of interest. In fact, the shear of a
filamentary structure is aligned orthogonally to its major axis
thereby defining the signal we are looking for.  We demonstrated that
this decomposition is optimal in maximizing the filament's signal when
combined with the optimal filter and allows its clear separation from
the central cluster isotropic contribution. This separation of
components derives from the fact that the filter signal for the cluster
is nearly constant for all PAs, being isotropic, while it
has an angular dependence for all anisotropic signal contributions
such as those produced by filaments. In particular, the signal peaks
when the filter is aligned with a filament. Note that the adopted
decomposition is substantially different from the one usually used,
implicitly or explicitly, in usual weak lensing mass reconstructions
or object detection and great care has to be taken when comparing and
interpreting different approaches.  With our proposed method, the
minimum convergence detectable in the filament's smooth component, at
$1\sigma$ level, lays in the range $0.005<\kappa_0<0.02$ depending on
the filter's template width and length. This performance is clearly
superior to other means of filament detection.

We stress that the filter is sensitive to all anisotropic signal
components such as a possible central cluster ellipticity which would
be degenerate with the presence of a filament. This is
not an issue, since the filter can be used in different ways 
depending on the question asked. In
particular: (1) By using all data, such that
the overall system anisotropic component (cluster ellipticity plus
filament) will be measured. One would expect the two components add
constructively since they are physically correlated; (2) By masking
out a circular area centered on the main central cluster such that the
filament's convergence alone is measured; (3) By running the filter
twice: The first time only on an area centered on the cluster and the
second time only on the rest of the field. The last option is of
particular interest because it allows to quantify the correlation
between the cluster ellipticity and the presence or the absence of a
filament, together with their respective alignment.

In addition, we present a simple method based on weak lensing peak
counts. Those sub-structure halos are expected to be more abundant
along the filament with respect to the rest of the field, contributing
to the individual identification of filaments. Also this second method
was tested on the N-body numerical simulation containing straight
filamentary structures. This approach complements the previous one and
helps to obtain a better picture of the amount of dark matter
contained in the filament's both, smooth and collapsed component.

Despite the simplicity of the proposed two methods we have proven
their efficiency to detect the dark matter content of intra-cluster filaments
in Mock observations. This statement holds for ground-based data quality
parameters, defining new possibilities in the study of structure
formation and cosmology via weak gravitational lensing.

\acknowledgements{We are grateful to Klaus Dolag for providing us the
  N-Body numerical simulation used in this work and to Massimo Meneghetti for his help 
  in extracting its lensing properties. 
  This work was supported by the Transregional
  Collaborative Research Centre TRR~33 (MM). This research was carried
  out in part at the Jet Propulsion Laboratory, California Institute
  of Technology, under a contract with NASA.  }

\bibliographystyle{aa}

\end{document}